# Quantifying Graft Detachment after Descemet's Membrane Endothelial Keratoplasty with Deep Convolutional Neural Networks

Friso G. Heslinga[1,*], Mark Alberti[2,*], Josien P.W. Pluim[1], Javier Cabrerizo[2,3,**], Mitko Veta[1,**]
1. Department of Biomedical Engineering, Eindhoven University of Technology, Eindhoven, the Netherlands
2. Ophthalmology Department, Rigshospitalet - Glostrup, Copenhagen, Denmark
3. Copenhagen Eye Foundation, Copenhagen, Denmark
*, ** contributed equally

**ABSTRACT**

**Purpose:**  We developed a method to automatically locate and quantify graft detachment after Descemet's Membrane Endothelial Keratoplasty (DMEK) in Anterior Segment Optical Coherence Tomography (AS-OCT) scans.

**Methods:**  1280 AS-OCT B-scans were annotated by a DMEK expert. Using the annotations, a deep learning pipeline was developed to localize scleral spur, center the AS-OCT B-scans and segment the detached graft sections. Detachment segmentation model performance was evaluated per B-scan by comparing (1) length of detachment and (2) horizontal projection of the detached sections with the expert annotations. Horizontal projections were used to construct graft detachment maps. All final evaluations were done on a test set that was set apart during training of the models. A second DMEK expert annotated the test set to determine inter-rater performance.

**Results:**  Mean scleral spur localization error was 0.155 mm, whereas the inter-rater difference was 0.090 mm. The estimated graft detachment lengths were in 69% of the cases within a 10-pixel (~150μm) difference from the ground truth (77% for the second DMEK expert). Dice scores for the horizontal projections of all B-scans with detachments were 0.896 and 0.880 for our model and the second DMEK expert respectively.

**Conclusion:**  Our deep learning model can be used to automatically and instantly localize graft detachment in AS-OCT B-scans. Horizontal detachment projections can be determined with the same accuracy as a human DMEK expert, allowing for the construction of accurate graft detachment maps.

**Translational Relevance:**  Automated localization and quantification of graft detachment can support DMEK research and standardize clinical decision making.

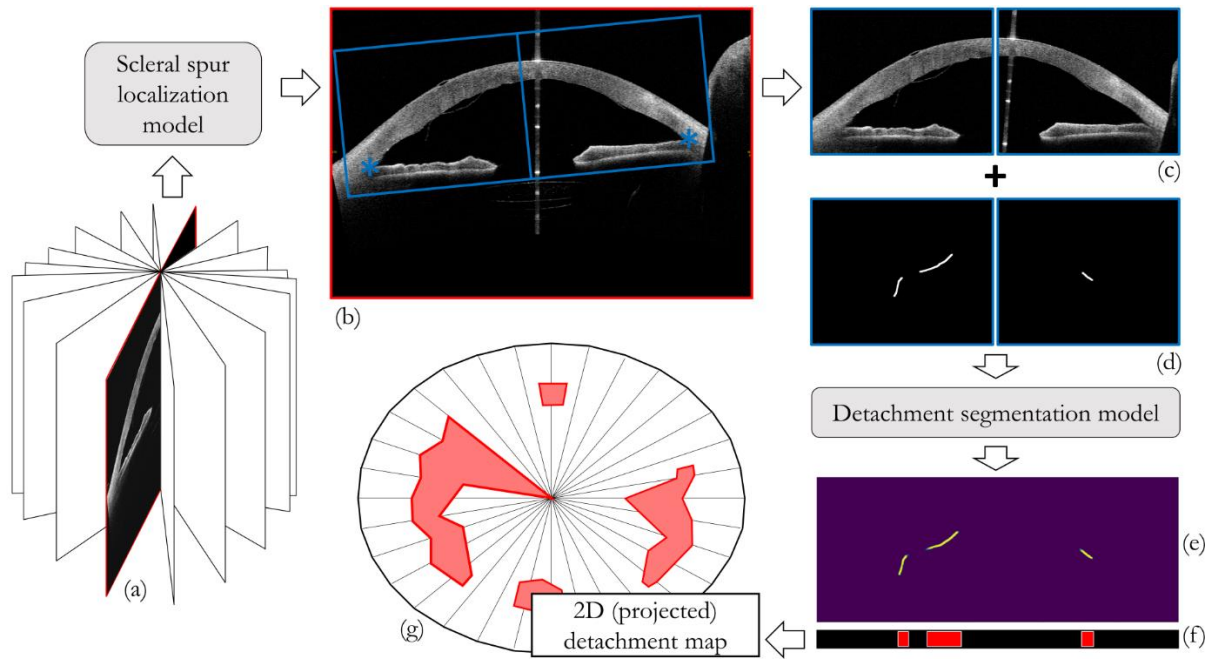

**Figure 1.** Deep learning pipeline for quantification of corneal graft detachment. A scleral spur localization model is applied to a radial B-scan of an AS-OCT (a). The scleral spur estimates are used to center the B-scan (b) and obtain crops. The crops (c) are processed through a segmentation model, which was trained to output a map with detachment predictions (e) similar to expert annotations (d). Combining the horizontal projections (f) of 16 B-scans, a graft detachment map can be constructed.

## 1. INTRODUCTION

Descemet's Membrane Endothelial Keratoplasty (DMEK) currently offers the greatest opportunity of visual gain to patients suffering from endothelial dysfunction.[1,2] However, partial graft detachment after DMEK remains a burden for patients and a challenge for surgeons with detachments requiring air injection in 3% to 76% of cases.[3,4]

Anterior Segment Optical Coherence Tomography (AS-OCT) allows for visualization of early graft detachment and is therefore clinically useful in guiding postoperative care.[5] However, the quantification of the detached area remains difficult because AS-OCT typically consists of multiple radial B-scans and a physician must integrate several images to have an overview of detached areas. Moreover, we found that the degree of graft detachment can be ambiguous in some regions when the graft is appositioned to the inner cornea yet not attached. No fast and objective tool to visualize all detached areas currently exists. We believe such a tool could aid in the postoperative management of DMEK patients, including the decision to rebubble or perform re-DMEK.

We propose an automated image analysis method (Figure 1) that has the potential to improve clinical decision making by objectively detecting the areas of DMEK detachment and providing an overview of all detached areas at once.

## 2. METHODS

**Ethical approval**

This study was approved by the Capital Region Committee on Health Research Ethics, Denmark and adhered to the tenets of the Declaration of Helsinki. All participants provided written informed consent before participation.

**Data**

Swept-source AS-OCT scans *(CASIA2; Tomey Corp. Nagoya, Japan)* were collected as part of a randomized controlled trial conducted at the Department of Ophthalmology, Rigshospitalet – Glostrup, Denmark. Briefly, the randomized study included patients with Fuchs' endothelial dystrophy or pseudophakic bullous keratopathy eligible for DMEK surgery and excluded re-DMEK procedures or prior keratoplasty. The study was double-blinded and was designed to compare patients randomized to either air or sulfur hexafluoride (SF6) DMEK surgery.[6] A DMEK expert (M.A.) annotated 80 scans from 68 participants, acquired either immediately after surgery and/or postoperative day 7. Typically, due to the present of a large gas bubble supporting most of the graft, little to no detachment is present immediately after surgery. However, these scans were included to help our model distinguish between graft detachment and intraocular gas. Each scan consists of 16 radial B-scans, corresponding to a total of 1280 images of 2133 by

1466 pixels. For each B-scan, locations where the graft had detached were manually annotated with point markings (image coordinates). Additionally, the scleral spur was annotated when clearly discernible in the inferior and superior part of the B-scan, resulting in a maximum of two points per scan. The data were randomly split on a participant level in a set for training and evaluation of our models (*n* = 960 images) and a set for final testing (*n* = 320 images). Details about the AS-OCT B-scans are shown in *Table 1*. Participant characteristics are shown in *Table 2*.

Table 1. Data sets details

|  | Training / validation data | Test data | Total |
|---|---|---|---|
| Participants | 50 | 18 | 68 |
| Hospital visits | 60 | 20 | 80 |
| AS-OCT B-scans | 960 | 320 | 1280 |
| AS-OCT B-scans without graft detachment | 232 (24.2%) | 104 (32.5%) | 336 (26.3%) |
| Scleral spur points annotated | 847 (44.1%) | 276* (43.1%) | 1123 (43.9%) |
| Both scleral spur points annotated in an AS-OCT B-scan | 288 (30.0%) | 81 (25.3%) | 369 (28.8%) |

Overview of image data and annotations. Data in the training and validation column was used to design, train and optimize the deep learning models. Test data was used for the final model evaluation. *out of 276 scleral spur points annotated by the first DMEK expert, 232 points were also annotated by a second DMEK expert to evaluate inter-rater agreement.

Table 2. Participant characteristics

| Characteristics | Participants in training and validation set | Participants in test set | p-value |
|---|---|---|---|
| Age, mean years (SD) | 70.2 (7.47) | 73.3 (6.10) | 0.09 |
| Sex (male/female) | 27 / 23 | 6 / 12 | 0.17 |
| Diagnosis (FED/PBK) | 50 / 0 | 17 / 1 | 0.26 |
| Laterality (R/L) | 26 / 24 | 10 / 8 | 1.00 |
| Tamponade (air/SF6) | 28 / 22 | 9 / 9 | 0.78 |
| Visit (immediately after surgery / postoperative day 7) | 10 / 50 | 2 / 18 | 0.72 |

Values are presented as mean (SD) or ratio. P-values were determined with T-test for continuous variables and Fisher's test for categorical variables. FED - Fuchs' endothelial dystrophy, PBK - pseudophakic bullous keratopathy.

**Deep learning pipeline**

For the analysis of the AS-OCT data, we used deep learning methodology,[7] which has successfully been used for many medical image analysis tasks[8] including ophthalmology.[9-12] In this paper we present a framework with a four-step approach: (1) localization of the scleral spur using a deep learning-based regression model to center each AS-OCT B-scan; (2) fit of an ellipse to the scleral spur points of all radial B-scans to refine localization and centering; (3) segmentation of the detached areas with a deep learning segmentation model and (4) extraction of DMEK biomarkers from the segmentation maps. Each step is described in more detail in the following sections. An overview of the deep learning pipeline is shown in Figure 1. Models were implemented in Keras,[13] using a TensorFlow backend.[14]

**Scleral spur localization**

For a clinical evaluation of graft detachment, and to study detachment progression, it is important to find the detached areas with respect to the center of the cornea. The center of the cornea is difficult to locate – especially when corneal edema is present, which is why we used the center of the fitted scleral spur ellipse instead. The scleral spur´s morphology and position have been proven to stay unaltered after surgery and therefore, it has been chosen as a landmark for quantitative measurements in the anterior chamber.[15,16] It is visible in only 70 - 78.9% of all of the radial B-scans,[17] due to image artifacts from eyelids or anatomical variations.[18] The localization model is therefore only trained on B-scans for which the scleral spur could be annotated in both the superior and inferior segment (n = 288). Training was done using batches of ten image crops and reducing the mean squared error. Basic data augmentation (rotation and translation) was used to increase the variability of the training data.[19]

AS-OCT B-scan images (2133 by 1466 pixels) were reduced to 512 by 352 pixels for and converted to grayscale. A well-known deep learning architecture, ResNet-50,[20] was modified to match the input dimensions and output four values that represent the coordinates of two scleral spur locations per B-scan. The localization model was trained with batches of 10 images by reducing the least-square error between the model outputs and the targeted coordinates. A grid search was used to find the optimal set of model hyper-parameters and select the best-performing model. This model was then used to process all B-scans in the test set. Note that scleral spurs locations were estimated even in B-scans that were not annotated.

The anatomical structure of the scleral spur can be approximated by an ellipse in the 3D AS-OCT volume. Since the scleral spur is not clearly discernible in each B-scan, we included an extra step that exploits this ellipsoid structure and makes the localization model robust for all B-scans. First, we fitted an ellipse through the 32 scleral spur point estimates (2 scleral spur points for each of the 16 AS-OCT B-scans). Then, for each scleral spur point estimate, we updated the estimate with the location of the ellipse through that slice.

**Detachment segmentation**

We created binary masks from the point markings that represent locations along the detached graft. Examples of the masks can be seen in Figure 4 and Figure 5. The width of the detached lines was set to 15 pixels. Based on the scleral spur point estimates of the scleral spur localization model, B-scans were cropped such that the cornea was centered (1920 by 768 pixels). Taking advantage of the anatomical symmetry of the anterior chamber, the crops were split into an inferior half and a vertically reflected superior half. This step halved the detachment model input size, while doubling the number of training examples. The crops were downsampled by a factor of two to obtain the final size of 480 by 384 pixels.

As a data augmentation technique, we added random uniform noise to the locations of the scleral spurs (-60 to +60 pixels in the horizontal and vertical coordinate) prior to cropping, resulting in translated and slightly rotated crops. The same cropping procedure and data augmentation was applied to the masks, ensuring that the OCT crops and masks remained aligned.

To localize the image pixels that illustrate graft detachment, we employed a semantic segmentation approach. A deep learning model with a U-Net architecture,[21] was implemented to output a mask similar to the input. The model was trained using batches of eight image crops, using a weighted cross-entropy loss. We experimented with the weight factor of foreground pixels on the loss, and found that a factor of 2 provided the best results on the validation set. The best performing model was applied to the test set to obtain mask predictions.

**Biomarkers extraction and evaluation**

For each mask prediction in the test set, *length of detachment* was determined using a skeletonization method.[22] This skeletonization method is a morphological procedure that involves shrinking the regions in the binary image until they are one pixel wide. The remaining pixels were counted as a proxy measure for length of detachment. After processing the test set crops with the detachment segmentation model, the skeletonization method was applied to the outputs of the segmentation model as well as the annotated masks. Evaluation of length of detachment was done by comparing the model predicted detachment length with the annotated length. Although length of detachment is our primary outcome measure, it does not provide information about the relative location of detachment. To enable the construction of a 2D-map of detachment, we projected the detached locations on the horizontal axis of each cropped radial B-scan. The 16 projections can then be combined to create a 2D-map giving an overview of all detached areas in a single image (Figure 1). Performance of the projection of the detached sections on the horizontal axis was evaluated using Dice score.[23] The Dice score was determined for the overlap between the projections and perfect overlap would result in a Dice score of 1.

Additionally, an inter-rater analysis was performed where a second DMEK expert (J.C.) annotated the B-scans in the test set. These annotations were processed similarly as the annotations of M.A. and assessed for scleral spur localization error, length of detachment and overlap in horizontal axis projection of detached graft sections.

## 3. RESULTS

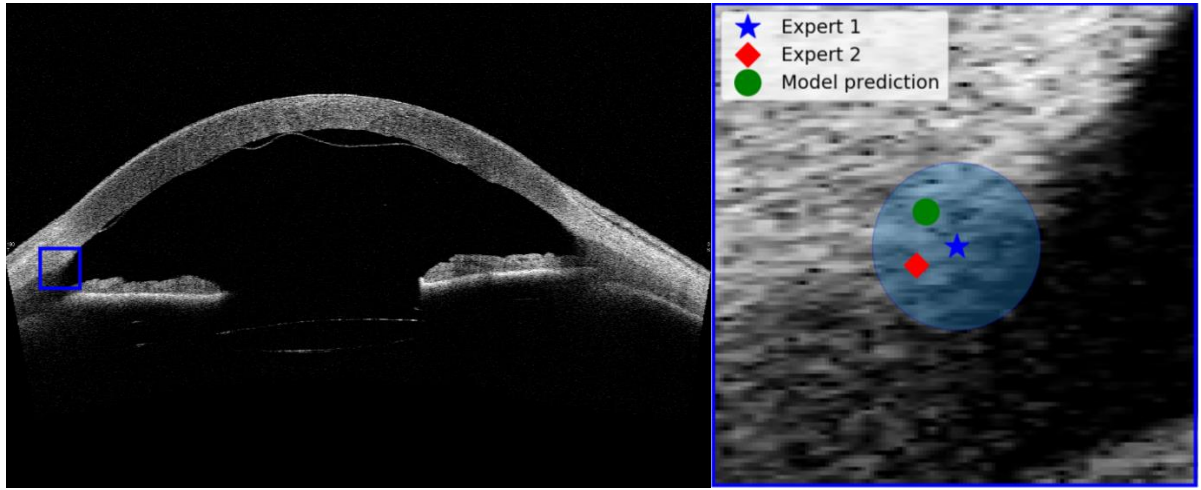

Figure 2. Example result of scleral spur localization model for a test case. Left: B-slice with original resolution (2133 x 1433 pixels). Right: Enlarged version of the blue box in the left image. The circle boundary represents the mean error between Expert 1 and the model prediction (0.155 mm).

**Scleral spur localization**

The scleral spur localization model with the ellipse fit was applied to the downscaled B-scans of the test set. The mean Euclidean distance between the annotations of Expert 1 and the model predictions was 4.97 pixels (0.155 mm). Moreover, 95% of these errors were within 8.79 pixels (0.275 mm). In comparison, the Euclidean distance between the two experts was 2.87 pixels (0.090 mm). Only scleral spur points that were annotated by both experts were used for the final evaluation. Figure 2 provides a visual interpretation where we show an example case and the mean error of the scleral spur localization model. We also tested for the effect of the ellipse fit. Without ellipse fit the mean Euclidean distance between the localization model and Expert 1's annotations was found to be slightly smaller: 4.48 pixels (0.140 mm). Additional discussion of our motivation to use the ellipse fit can be found in the discussion section.

**Length of graft detachment**

The main results of the segmentation model are shown in Figure 3, where length of detachment is displayed in a Bland-Altman plot.[24] The original field of view of the B-scan was 16 by 11 mm, for 2133 by 1466 pixels, so after downscaling with a factor two, one pixel corresponds to 15.0 μm. The bias (6.04 pixels) is relatively small compared to the mean length of detachment, and 69% of cases are within a difference of 10 pixels (~150μm). Some outliers are found for cases with a larger length of detachment and these mostly represent underestimations of the length of detachment. In comparison, the bias in detachment length between Expert 1 and Expert 2 was -0.9, with 1.96 SD (standard deviations) between -33.56 and 31.44. For 77% of cases, the difference in annotated length of detachment is within 10 pixels.

The numbered green dots in Figure 3 refer to specific B-scans that are shown in Figures 4 and 5. Green dots 1-3 are examples of successfully segmented scans, while numbers 4-6 correspond to B-scans for which the segmentation model outputs show substantial deviations from the expert annotations.

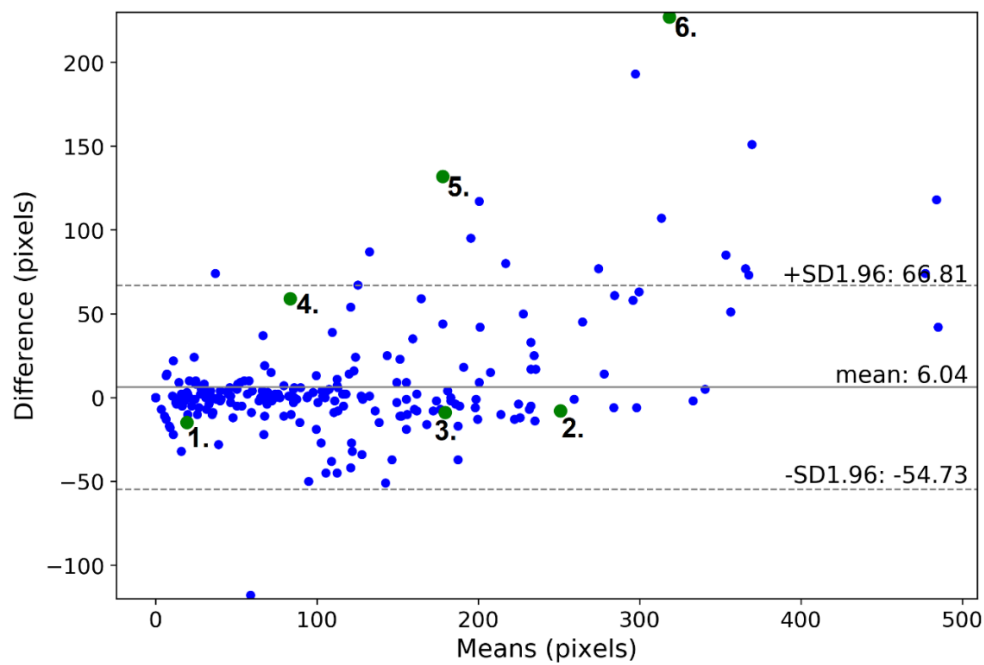

Figure 3. Bland-Altman plot of length of detachment determined by the segmentation model versus the annotations of Expert 1. Length is measured as the number of pixels after applying a skeletonization method to the mask. The horizontal axis describes the mean of the length of detachment as determined by Expert 1 and the segmentation model. The vertical axis is the difference between Expert 1 and the segmentation model. ±1.96 SD (standard deviations) describes the 95% confidence interval. A positive difference means that the segmentation model underestimates detachment length compared with the expert annotations. One pixel corresponds to 15.0 μm.

The test set included two OCT scans that were acquired immediately after surgery. In all 32 B-scans, the edge of the intraocular gas bubble was visible to the human observer, but no graft detachment was present. The detachment model provided false positive regions in 2 out of the 32 B-scans.

**Projection results**

When all B-slices were included, the Dice score was found to be 0.906 (±0.190), compared to 0.916 (±0.160) for the inter-rater performance. When empty masks were excluded from this analysis, the Dice scores for the segmentation model and the inter-rater performance were found to be 0.896 (±0.149) and 0.880 (±0.172) respectively.

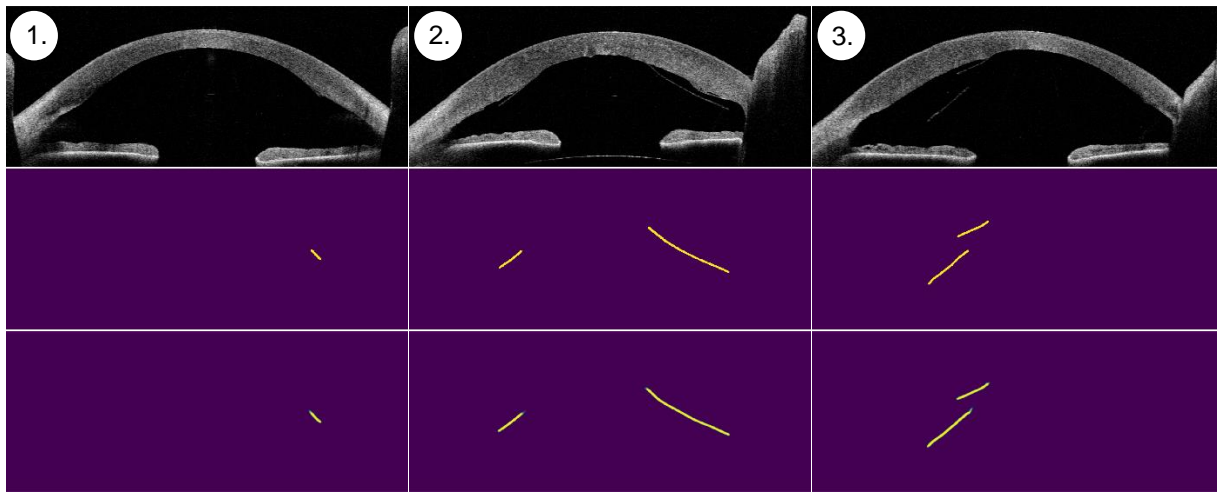

Figure 4. Examples of successful segmentations. Top row: OCT B-slices from the test set. Middle row: mask annotations by a DMEK expert. Bottom row: output of the segmentation model. For the predictions, yellow indicates high confidence that a section is detached, while green indicates lower confidence. The numbers in the top left corner correspond to the green dots in Figure 3.

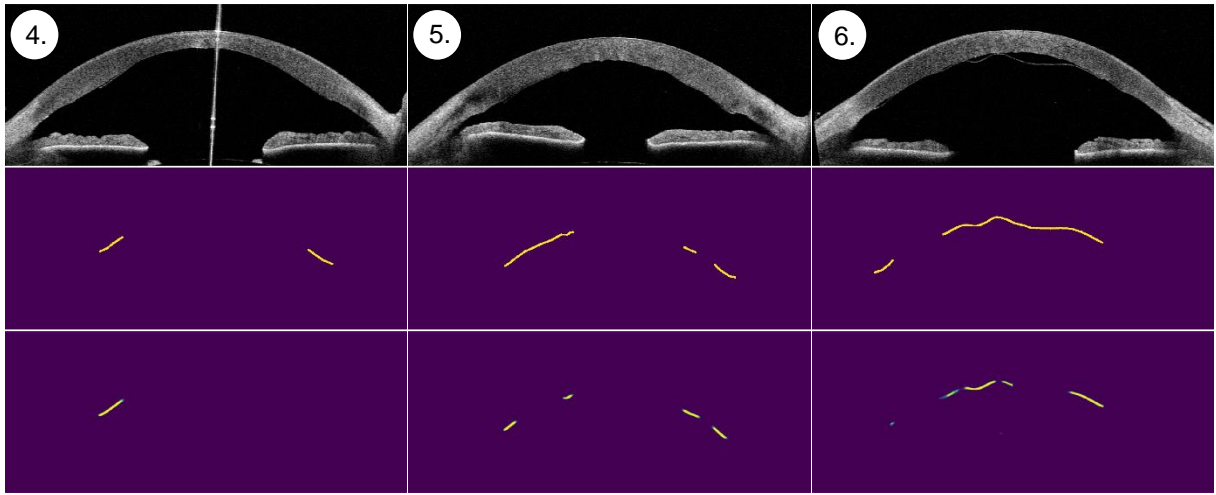

Figure 5. Examples of segmentations that deviate from the expert annotations. For more details, see the description of Figure 4.

The detachment projections of 16 B-scan can then be plotted on a grid similar to the radial grid of the AS-OCT scan. Since all B-scans were previously centered with respect to the middle of the cornea (using the scleral spur estimates), the detached sections can directly be mapped on the radial grid. Three examples of such detachment maps are shown in Figure 6, in which the red structures represent the detachment segmentation model estimates and the green dotted line the expert annotations on AS-OCT images.

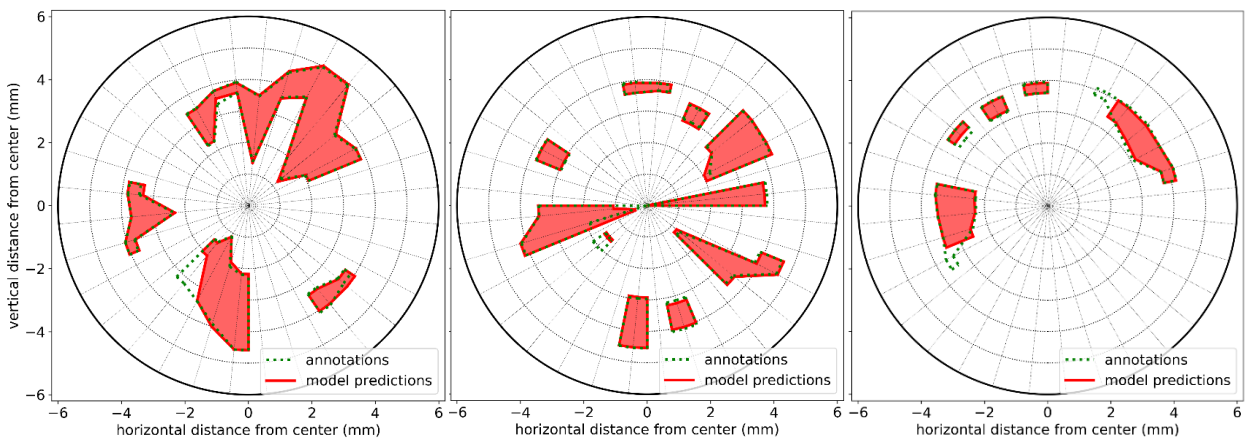

Figure 6. Graft detachment maps of three AS-OCT scans, connecting the detached sections of 16 radial B-scans. The red line and surface indicates the model predictions, while the green dotted line represents the expert annotations.

## 4. DISCUSSION

Our results demonstrate that a deep convolutional neural network can accurately and automatically identify DMEK graft detachment. We believe that our deep learning pipeline has the potential to improve and standardize clinical decision making and can similarly be used as an objective and operator-independent outcome to improve DMEK research and reporting.

The number of DMEK procedures performed is rising rapidly driven by the superior visual results.[25,26] In 2018, 41.2% more DMEKs were performed in the United States, while the total number of endothelial keratoplasty procedures increased only 4.6% in comparison with the year before.[27] This invariably increases the need for DMEK detachment management, such as the decision to await spontaneous clearance; rebubble; or perform re- DMEK.[28-30] Studies agree that management depends on the degree of detachment yet report diverging opinions for when to rebubble and varying definitions of partial detachment, including *visually significant graft detachment*,[31] *20% detached area*,[32] *more or less than 1⁄3 detached*.[28] In current practice, the amount of detachment after DMEK is estimated by a clinician/surgeon over a succession of scans on a screen, rather than measured objectively. Thus the surgeon has to make a decision with regards treatment without seeing all detached areas in a single image or accurately being able to quantify the total amount of detachment.

The high Dice scores for the projection results are similar to a human DMEK expert and indicate that accurate detachment maps can be constructed. Visual evaluation of some examples of these maps (Figure 6) indeed shows a strong similarity with the expert annotations. Since the center of the detachment map corresponds with the center of the cornea, the severity of the detached sections can be evaluated with their respective distance to the center. Moreover, follow-up OCT scans can be overlaid to assess detachment progression.

The Bland-Altman plot in Figure 3 also indicates that the segmentation model works well for most individual B-scans. Examples 1-3 in Figure 4 represent the results for the majority of the segmentations and show high segmentation accuracy, even when a graft is torn (example 2). For some cases, the predicted detachment length differed substantially from the expert annotations (Figure 5). Part of the disagreement could originate in the inherent uncertainty of some graft sections that are difficult to annotate. Indeed, the DMEK experts do not always agree, but the 95% confidence interval for the inter-rater study is roughly half the size of the model prediction confidence interval. Moreover, we also found that the model makes a few substantial mistakes that are obvious to the human observer (e.g. example 6). After visual inspection of the outliers, we noticed that most of the sizeable underestimations were B-slices of one specific OCT scan with a lot of detachment in the center. These errors are likely due to the lack of training examples with a large central detachment. The model might confuse some large center detachments for intraocular gas, which is only present in scans directly after surgery or rebubbling. Although the effect of these type of mistakes might be limited since they are obvious and will easily be spotted by the ophthalmologist, it could be addressed by adding more training data encompassing more variations, especially cases with center detachments. Furthermore, the current segmentation model did not take into account information from neighboring B-slices, as the DMEK expert did. Finally, some inaccuracy might result from the loss of information due to downsampling the B-slices by a factor two before processing the scans with the graft segmentation model. Given the horizontal line-like structure of the graft and the downsample factor, the horizontally most distant pixels could be misclassified. However, this error will be small compared to the whole length of the graft detachment.

Apart from missegmenting some cases with a lot of detachment in the center, it was sometimes challenging to distinguish remnant host Descemet's membrane from the DMEK graft. Furthermore, the presented models were trained and evaluated on a single data set collected with one type of AS-OCT device. For generalization towards multiple-sources, the models have to be either retrained with some images from other scanner types, or with the use of other domain generalization techniques.[33]

Prior image analysis work within the realm of DMEK detachment has only focused on binary classification; i.e. whether detachment is present or not[34] and whether rebubbling was performed.[35] We believe our detachment model is of clinical value as it provides quantitative measures about length and location of graft detachment. The segmentation accurately locates detachment in most AS-OCT B-scans and is much faster than a human rater. In clinical practice an ophthalmologist would not have time to annotate the detachment regions in detail, while our deep learning pipeline could provide an instant evaluation aiding the decision.

Although our aim was to develop a model for quantifying DMEK detachment, we also developed a scleral spur locating model as an intermediate step. Having this scleral spur localization model aided our AS-OCT B-scan preprocessing by cropping all images uniformly prior to the DMEK detachment model evaluation. This cropping step also provided practical benefits, as we did not have to reduce the standard U-net model size or the resolution of the B-scans further to fit within GPU memory. However, locating the scleral spur is valuable in and of itself. Potential applications include determining limbal chamber depth parameters such as angle-opening distance (AOD) and trabecular-iris space area (TISA), relevant in glaucoma. Furthermore, it may also be a valuable tool for aligning AS-OCT scans between patient visits (e.g. to compare pachymetry map changes).

The refinement of the scleral spur estimates by fitting an ellipse resulted in a slightly bigger localization error. However, we could only evaluate for scleral spur points that were well discernable, since those were annotated by both experts. Our model also outputs an estimate for the scleral spur when the region itself is not visible (e.g. hidden behind the eyelid).[13] We think that the ellipse fit step makes the localization more robust for these cases and reduces outliers. Whether our scleral spur model can be applied to other disease entities, such as acute angle-closure glaucoma is a topic of future research.

In summary, we have introduced a deep learning pipeline based on AS-OCT that allows automatic and accurate quantification of graft detachment after DMEK. Our future research efforts will focus on evaluating the value of our algorithm for improving clinical decision making and clinical outcomes after DMEK.